\documentclass[aps,prl,twocolumn,superscriptaddress]{revtex4-2}
\usepackage[T1]{fontenc}
\usepackage[utf8]{inputenc}
\usepackage{amsmath}
\usepackage{amsfonts}
\makeatletter
\usepackage{xcolor}
\usepackage{soul}
\usepackage{comment}
\usepackage[english]{babel}
\setstcolor{red}
\usepackage[version=4]{mhchem}
\usepackage{bm}
\usepackage{graphicx}
\usepackage{indentfirst}
\makeatother

\begin{document}

\title{Local Density of States as a Probe of Multifractality in Quasiperiodic Moir\'e Materials
}

\date{\today}

\author{Ricardo Oliveira}
\thanks{These authors contributed equally to this work.}

\affiliation{Centro de Física das Universidades do Minho e Porto, LaPMET, Departamento de Física e
Astronomia, Faculdade de Ciências, Universidade do Porto, Rua do Campo Alegre s/n,
4169-007 Porto, Portugal}

\author{Nicolau Sobrosa}
\thanks{These authors contributed equally to this work.}

\affiliation{Centro de Física das Universidades do Minho e Porto, LaPMET, Departamento de Física e
Astronomia, Faculdade de Ciências, Universidade do Porto, Rua do Campo Alegre s/n,
4169-007 Porto, Portugal}
\author{Pedro Ribeiro}
\affiliation{CeFEMA, Instituto Superior Técnico, Universidade de Lisboa, Av. Rovisco Pais, 1049-001
Lisboa, Portugal}
\author{Bruno Amorim}
\author{Eduardo \surname{V. Castro}}
\affiliation{Centro de Física das Universidades do Minho e Porto, LaPMET, Departamento de Física e
Astronomia, Faculdade de Ciências, Universidade do Porto, Rua do Campo Alegre s/n,
4169-007 Porto, Portugal}

\begin{abstract}
Quasiperiodic moir\'e materials provide a new platform for realizing critical electronic states, yet a direct and experimentally practical method to characterize this criticality has been lacking. We show that a multifractal analysis of the local density of states (LDOS), accessible via scanning tunneling microscopy, offers an unambiguous signature of criticality from a single experimental sample. Applying this approach to a one-dimensional quasiperiodic model, a stringent test case due to its fractal energy spectrum, we find a clear distinction between the broad singularity spectra $f(\alpha)$ of critical states and the point-like spectra of extended states. We further demonstrate that these multifractal signatures remain robust over a wide range of energy broadenings relevant to experiments. Our results establish a model-independent, experimentally feasible framework for identifying and probing multifractality in the growing family of quasiperiodic and moir\'e materials.
\end{abstract}

\maketitle

Quasiperiodic systems are known for their unconventional electronic
structure and critical multifractal states, having recently found
a new experimental platform in moir\'e materials, whose incommensurability arises from the
misalignment or twist between the layers. In twisted bilayer
graphene (tBLG), localization effects has been predicted for incommensurate
twist angles, including in the so-called dodecagonal graphene quasicrystal
at $30\text{º}$, \cite{PhysRevB.99.165430,PhysRevB.99.245401}. 
Multifractal momentum-space wavefunctions were shown to emerge  near
the magic-angle semimetal critical regime in the chiral limit at intermediate
angles $\sim9\text{º}$ \cite{Fu2020} and sub-ballistic states were also shown to be
delocalized in momentum-space within the narrow-band region around the first magic-angle
\cite{Goncalves2021}. Another striking realization is found in twisted
trilayer graphene, where twisting three layers of graphene at two
different angles creates incommensurate moir\'e patterns that form a
tunable moir\'e quasicrystal \cite{Uri2023,PhysRevLett.133.196401,Hao2024}, where new forms of superconductivity are only explained through the lens of quasiperiodicity.
Furthermore, heterostructures of tBLG stacked on hexagonal boron nitride
(hBN) provide another route to tunable quasiperiodic structures, encompassing
quasicrystals with Bravais-forbidden dodecagonal symmetry as well
as intercrystals that lack forbidden symmetries \cite{Lai2025}. These discoveries highlight the central role of critical wavefunctions in understanding quasiperiodic moir\'e systems.

Wavefunctions at criticality are known to display complex spatial
fluctuations on many length scales. A powerful theoretical framework
for studying such complexity is multifractal analysis, originally
introduced to characterize the scale-invariant structure of measures
in turbulence and dynamic systems \cite{PhysRevA.33.1141},
but later entered the condensed matter community in the context of wavefunction criticality in disordered electronic systems \cite{JANSSEN1994,RevModPhys.80.1355}.
At metal-insulator transitions, wavefunctions are characterized by a continuum of fractal dimensions, forming a broad singularity spectrum {$f(\alpha)$}. Multifractality has been extensively studied
in disordered systems, including the 3D Anderson transition \cite{PhysRevLett.67.607,PhysRevB.62.7920,PhysRevLett.97.046803,PhysRevLett.105.046403,PhysRevB.84.134209}
and the quantum Hall effect \cite{RevModPhys.67.357,PhysRevB.64.241303}. 

This methodology has since evolved beyond its theoretical origins
and has been employed in experimental data from scanning
tunneling microscopy (STM) in disordered
systems. For instance, it has been applied to the study of the metal-insulator transition of cleaved \ce{Ga_{1-x}Mn_{x}As} samples \cite{Richardella2010}, the superconducting phase of
weakly disordered single-layer \ce{NbSe_{2}} \cite{rubio2020visualization},
\ce{Bi_{x}Pb_{1-x}/Ag(111)} surface alloys \cite{Jaeck2021}, structurally
disordered \ce{MoS_{2}} monolayer semiconductors \cite{PhysRevResearch.5.043029},
disordered nodal line semimetals \ce{Fe_{3}GeTe_{2}} \cite{mathimalar2025concurrent}
and monolayers of tin on silicon \cite{lizee2025anderson}. Across all of these studies, differential conductance (dI/dV) maps provide a spatially
resolved measurement of the local density of states (LDOS), which
encodes signatures of the underlying eigenstates. When systems approach
criticality, the LDOS itself can become multifractal, a feature which can be unveiled through an analysis of dI/dV data.

Despite its success in disordered systems, multifractal analysis has
not yet been widely adopted in the study of quasiperiodic systems,
particularly in moir\'e materials. This is surprising given that many
quasiperiodic systems display critical wavefunctions exhibiting deterministic
multifractality \cite{Evangelou1987, Siebesma1987, 1992IJMPB...6..281H}. Nevertheless, multifractal analysis has been used to study the Fibonacci chain \cite{PhysRevB.93.205153} as well as the two-dimensional Penrose and Ammann–Beenker tilings \cite{PhysRevB.96.045138}. More recently, it has also been applied to compute the singularity spectrum of resonance amplitudes in chains of small high-index dielectric resonators \cite{PhysRevB.108.064210}, providing an experimental realization of the Fibonacci chain.

In this Letter, we introduce a practical and experimentally accessible
approach to identify clear signatures of quasiperiodicity using multifractal
analysis of the LDOS, directly measurable through differential conductance maps obtained from STM. Unlike conventional finite-size scaling
methods commonly employed in theoretical studies of quasiperiodic
systems, our approach relies on the multifractal canonical partition
function, thereby eliminating the need for data across multiple sample
sizes. Crucially, our analysis reveals
quantitatively how multifractality emerges in the LDOS, providing
an unambiguous distinction between spectral regions characterized
by extended Bloch states and those dominated by quasiperiodic effects.
This method offers experimentalists an immediate tool for probing and understanding multifractal signatures in real quasiperiodic materials.
\begin{figure}[h!]
    \centering
    \includegraphics[width=0.95\linewidth]{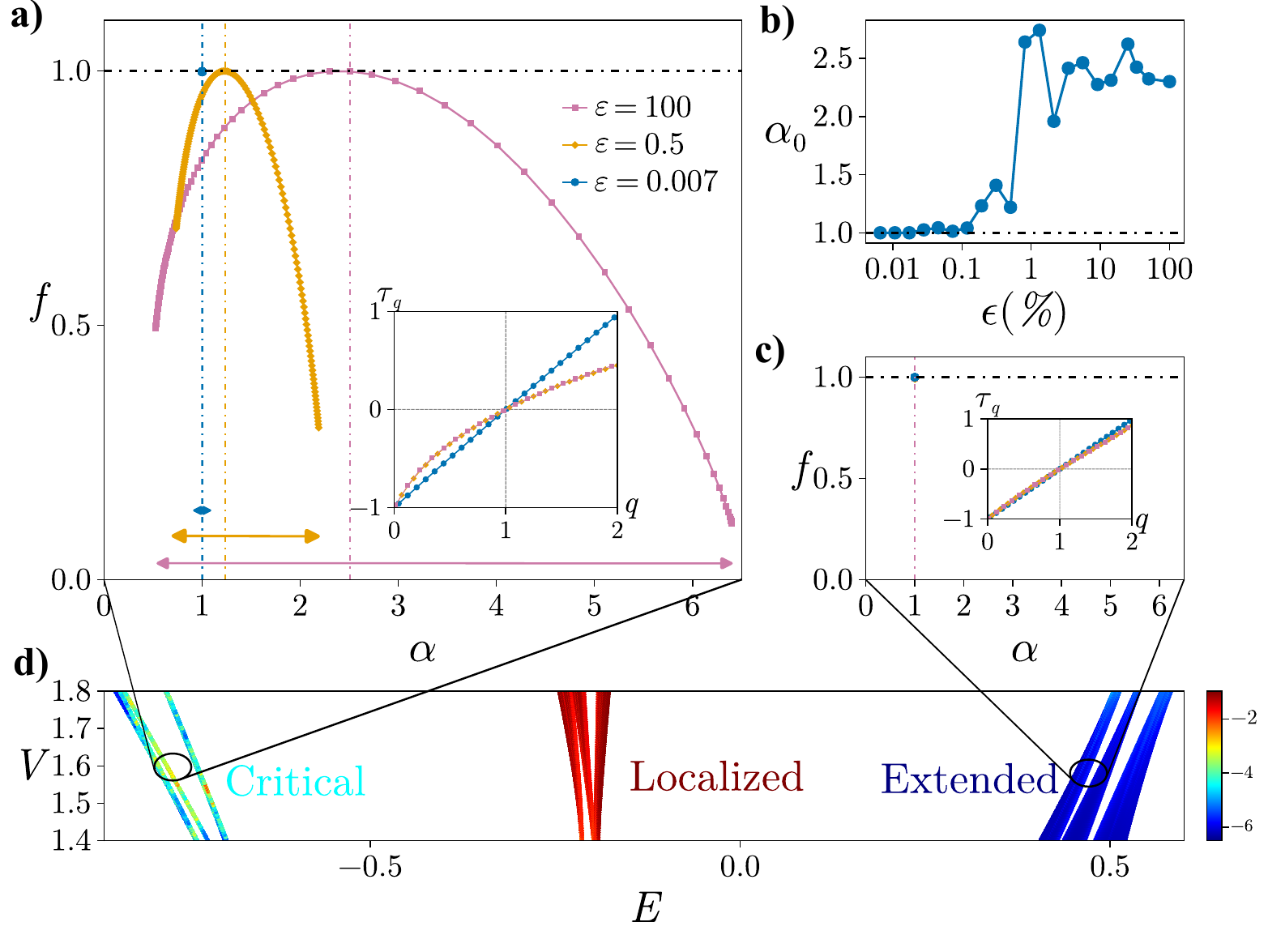}
    \caption{
    (a) Singularity spectrum $ f(\alpha) $ of the LDOS at a fixed energy in the critical narrow band for different quasiperiodic approximants with total system sizes $ L \simeq 1.2\times10^{5} $. Increasing $\varepsilon$ (larger unit cell size) enhances multifractality as we approach the quasiperiodic limit ($\varepsilon = 100\%$). Inset: corresponding nonlinear Legendre transform $ \tau_{q} $.
    (b) Position of the spectral maxima $ \alpha_{0} $ vs $ \varepsilon$; $ \alpha_{0}>1 $ in the quasiperiodic limit and $ \alpha_{0}\approx1 $ for Bloch states.
    (c) Same as in (a) for extended states, yielding point-like singularity spectra and linear $ \tau_{q} $.
    (d) Energy spectrum vs quasiperiodic modulation \( V \); colors denote the inverse participation ratio, distinguishing the extended, localized, and critical narrow bands.
    }
    \label{fig:Singularity spectra}
\end{figure}

Fig.\,\ref{fig:Singularity spectra} illustrates the multifractal analysis of the LDOS as a method for detecting quasiperiodicity. Panel (a) shows the singularity spectrum $f(\alpha)$, for the LDOS at energies hosting critical states. The width and shape of the spectra provide clear signatures of quasiperiodic behavior. Each curve corresponds to a different rational approximant, flowing from the periodic limit (point spectra) toward the quasiperiodic limit (broad spectra) where critical states emerge. In contrast, panel (c) presents the same analysis for extended states, where all curves collapse into narrow, nearly point-like spectra, a hallmark of conventional metallic behavior. These distinct spectral profiles offer a practical criterion for distinguishing between critical and extended states in quasiperiodic systems.

\paragraph{Model}To demonstrate how the multifractal analysis of the LDOS reveals energies at which quasiperiodicity influences
electronic properties, we study a representative 1D quasiperiodic
model whose phase diagram was explored by \citet{PhysRevLett.131.186303}.
This model hosts critical phases, coexisting with localized and extended
ones, exhibiting rich mobility edges, i.e., energy-dependent transitions
between these phases. Its Hamiltonian is given by 
\begin{align}
H & =t\sum_{n=0}^{L-1}\sum_{m=1}^{R}e^{i\alpha m-pm}c_{n}^{\dagger}c_{n-m}+h.c.\nonumber \\
 & +2V\sum_{n=0}^{L-1}\sum_{l=1}^{R}e^{-ql}\cos\left(2\pi\tau n\right)c_{n}^{\dagger}c_{n}\label{eq:Hamiltonian of our model}
\end{align}
where the first term represents exponentially decaying hopping amplitudes
with decay rate $p$ and a distance cutoff $R$, pierced by a magnetic
flux $\alpha/2\pi$ . The second term describes a quasiperiodic potential
constructed from harmonics of an incommensurate wavenumber $2\pi\tau$, $\tau \notin \mathbb{Q}$,
also decaying exponentially with decay rate $q$ and harmonic cutoff
$R$. Throughout this Letter, we assume periodic boundary conditions (PBC)
and fix parameters to $\left(p,q,\alpha\right)=\left(1,1.5,\frac{\pi}{2}\right)$. We show the energy spectrum as well as the localization properties as a function of the potential strength, $V$, in Fig.\,\ref{fig:Singularity spectra} (d), where we distinguish between critical, localized and extended states forming narrow bands of energy, with well-defined mobility edges.

In this work, we fix the incommensurate wavenumber to be the inverse
of the golden ratio $\tau=\frac{\sqrt{5}-1}{2}$. This
irrational number can be approximated by a sequence of rational approximants,
$\tau_{n}=\frac{F_{n-1}}{F_{n}}$, where $F_{n}$ denotes the $n$-th
Fibonacci number. To model systems that approach the quasiperiodic
limit, we consider periodic systems with sizes determined by the Fibonacci
sequence. Specifically, we construct Hamiltonians with PBC using unit
cells of size $L_{\text{UC}}=F_{n}$. The total system then consists
of $N$ such unit cells with size $L=NL_{\text{UC}}$, where $L$ is kept fixed for all considered systems.  In this
framework, periodic systems are constructed with low-order rational
approximants to $\tau$, and the quasiperiodic limit is reached as
$\tau_{n}\rightarrow\tau$. By properly choosing unit cell
sizes from the Fibonacci sequence, we can systematically study
and compare periodic and quasiperiodic systems of comparable total
size $L$. Since every system has a similar size, $L$, we are decoupling the quasiperiodicity effects from the finite-size ones. To quantify how close a given approximant is to the quasiperiodic limit, we introduce the parameter $\varepsilon \, (\%) \equiv \frac{L_{\text{UC}}}{L} \times100$. When $\varepsilon = 100
\%$, the system consists of a single unit cell and corresponds to the quasiperiodic case for our fixed total system size. As $\varepsilon \rightarrow 0$, the unit cell becomes small compared to the total system size, and we say the system approaches the periodic limit.

\paragraph{Multifractal Analysis} We now introduce the canonical partition function formalism for multifractal
analysis, originally developed in \cite{PhysRevLett.62.1327}, which
offers an alternative to traditional
finite-size scaling. This method is tailor-made for fixed system sizes
and is ideal for analyzing any spatially resolved quantities, in particular, the
LDOS. By
probing the inner structure of such observables, this approach allows
us to extract multifractal properties without needing data from multiple
system sizes. 

The method proceeds by partitioning the lattice into non-overlapping
boxes of size $l^{d}$, where $l$ is a fixed spatial resolution chosen
to divide the linear system size $L$, and $d$ the dimensionality of the problem. We define the coarse-graining scale as
$\lambda\equiv\frac{l}{L}$.
After square normalizing the LDOS across the lattice, $\sum_{\boldsymbol{R}}\left|\rho_{\boldsymbol{R}}\left(E\right)\right|^{2}=1$, to simplify the following definitions, we coarse-grain the LDOS into boxes of size $\left(\lambda L\right)^{d}$ to
define the probability measure 
\begin{equation}
P_{\boldsymbol{b}}\left(\lambda;E\right)=\sum_{\boldsymbol{R}\in B_{\lambda}\left(\boldsymbol{b}\right)}\left|\rho_{\boldsymbol{R}}\left(E\right)\right|^{2}, \label{eq:Probability measure}
\end{equation}
where $B_{\lambda}\left(\boldsymbol{b}\right)$ denotes the small partition, or "box", whose left corner is located at position $\boldsymbol{b}$ on the lattice. In other words, each $\boldsymbol{b}$ identifies the reference corner of one of the non-overlapping boxes used to coarse-grain the system. The sum in Eq.\,\eqref{eq:Probability measure} runs over all lattice sites $\boldsymbol{R}$ that lie within the box $B_{\lambda}\left(\boldsymbol{b}\right)$. This method can be extended to any scalar quantity defined on the lattice or on a sampling grid, such as, for example, the spatially-resolved superconducting gap \cite{rubio2020visualization}. The multifractal partition function at
coarse-graining scale $\lambda$ is then defined as 
\begin{equation}
Z_{q}\left(\lambda;E\right)=\sum_{\boldsymbol{b}}\left[P_{\boldsymbol{b}}\left(\lambda;E\right)\right]^{q},\label{eq:Multifractal partition function}
\end{equation}
which is essentially the generalization of the inverse participation ratio
(IPR) of the coarse-grained LDOS, that is, the $q$-th moment of the
LDOS.

If $Z_{q}\left(\lambda;E\right)$ exhibits scale-invariant behavior
over a range $\left[\lambda_{1},\lambda_{2}\right]$, then it obeys
a power-law of the form
\begin{equation}
Z_{q}\left(\lambda;E\right)\sim\lambda^{\tau_{q}\left(E\right)},\label{eq:Power-law-scaling-for-Z}
\end{equation}
allowing us to extract the multifractal exponent $\tau_{q}\left(E\right)$
directly as the slope on a log-log plot of $Z_{q}$ versus $\lambda.$

To fully characterize the multifractal properties of the LDOS, we
compute the singularity spectrum $f\left(\alpha\right)$, which describes
the fractal dimension of subsets of the system where the LDOS scales
locally with exponent $\alpha$. This spectrum is related to $\tau_{q}$
through the Legendre transform 
\begin{equation}
\alpha_{q}=\frac{d\tau_{q}}{dq},\quad f\left(\alpha_{q}\right)=q\alpha_{q}-\tau_{q}.\label{eq:Legendre-transform}
\end{equation}
The singularity spectrum can be obtained directly using the canonical
approach, described as follows: we define the normalized measure over boxes as 
\[
\mu_{\boldsymbol{b}}\left(q,\lambda;E\right)\equiv\frac{\left[P_{\boldsymbol{b}}\left(\lambda;E\right)\right]^{q}}{Z_{q}\left(\lambda;E\right)},
\]
which acts as a probability distribution. From this, both the multifractal
entropy 
\begin{equation}
S_{q}\left(\lambda;E\right)\equiv\sum_{\boldsymbol{b}}\mu_{\boldsymbol{b}}\left(q,\lambda;E\right)\log\mu_{\boldsymbol{b}}\left(q,\lambda;E\right)\label{eq:Multifractal-entropy}
\end{equation}
and the derivative of the partition function with respect to $q$
\begin{equation}
Z_{q}'\left(\lambda;E\right)\equiv\sum_{\boldsymbol{b}}\left[P_{\boldsymbol{b}}\left(\lambda;E\right)\right]^{q}\log P_{\boldsymbol{b}}\left(\lambda;E\right),\label{eq:Derivative of partition function}
\end{equation}
allow us to estimate the multifractal quantities via:
\begin{equation}
\frac{Z_{q}'\left(\lambda;E\right)}{Z_{q}\left(\lambda;E\right)}\sim\lambda^{\alpha_{q}\left(E\right)}\quad S_{q}\left(\lambda;E\right)\sim\lambda^{f_{q}\left(E\right)}.\label{eq:Alpha-and-f}
\end{equation}

Our primary local observable is the LDOS. To compute it, we use the
implicitly restarted Arnoldi method to extract the eigenvectors $\left\{ \phi_{\boldsymbol{R}}\left(E_{\mu}\right)\right\} $
and eigenvalues $\left\{ E_{\mu}\right\} $ close to the target energy
$E$. We then apply a Gaussian broadening to simulate finite spectral
resolution, controlled by a width $\eta$. The LDOS is thus defined
as
\begin{equation}
\rho_{\boldsymbol{R}}\left(E\right)=\frac{1}{\sqrt{2\pi\eta^{2}}}\sum_{\mu}\left|\phi_{\boldsymbol{R}}\left(E_{\mu}\right)\right|^{2}\exp\left(-\frac{1}{2}\frac{\left(E-E_{\mu}\right)^{2}}{\eta^{2}}\right).\label{eq:LDOS}
\end{equation}
We
aim to distinguish extended and critical phases by studying the multifractal
properties of the LDOS across different regions of the spectrum and
for systems with different numbers of unit cells.

To study multifractality, we can examine the multifractal exponent
$\tau_{q}$. For purely extended states, $\tau_{q}=d\left(q-1\right)$. In contrast, multifractal
systems are characterized by a nonlinear dependence of $\tau_{q}$
on $q$. A more insightful characterization of multifractality is
given by the singularity spectrum $f\left(\alpha\right)$. Extended states yield a sharp, single-point
spectrum $\left(\alpha,f\right)=\left(d,d\right)$, for every $q$, while multifractal
states display a broad, concave $f\left(\alpha\right)$ curve, spanning
from $\alpha_{\text{min}}$ to $\alpha_{\text{max}}$. The spectral
width $\Delta\alpha=\alpha_{\text{max}}-\alpha_{\text{min}}$ and the typical Hölder exponent $\alpha_{0}$, the position of the maximum of the spectrum, provide direct measures of multifractality. 

\paragraph{Results} 
Fig.\,\ref{fig:Singularity spectra} showcases the multifractal
properties of the LDOS in our quasiperiodic model. For the energy
narrow bands associated with critical states, we observe in panel
(a) that the singularity spectrum is broad. As
we increase $\varepsilon$, which corresponds to increasing the size of unit cell for a fixed total system size, and approach the quasiperiodic limit ($\varepsilon=100\%$) multifractality becomes more pronounced, as evidenced by the broader spectra, where the horizontal colored arrows mark the spectral width, $\Delta\alpha$ and the vertical dashed lines mark the typical value, $\alpha_0$. The inset also demonstrates the non-linear behavior of $\tau_{q}$, a quantity typically extracted from finite-size scaling analysis, but computed here with the partition function method. In panel (b) we show the position of the maximum of each spectra as a function of $\varepsilon$. In the quasiperiodic limit, the spectra are centered at $\alpha_0 >1$, a signature of multifractality. Going to the periodic limit, by decreasing $\varepsilon$ (i.e., reducing the unit cell size), the value oscillates and then stabilizes at $\alpha_0=1$, indicating the Bloch nature of the LDOS in this limit. This behavior contrasts with the results of panel (c), where for energies in the narrow bands associated with extended states, the LDOS yields sharp,
point-like singularity spectra, for all unit cell sizes, as expected. In the inset, the standard result emerges, with a linear $\tau_{q}$ for any approximant.

The clear emergence of nonlinear $\tau_{q}$ and broad $f\left(\alpha\right)$
curves is unambiguous evidence of multifractality in the LDOS of quasiperiodic
systems. Importantly, these signatures are found without the need
for finite-size scaling, due to the canonical partition function method, previously introduced. As such, in an experiment, a single measurement of the LDOS would be sufficient to detect multifractality.

\begin{figure}[h!]
    \centering
    \includegraphics[width=0.95\linewidth]{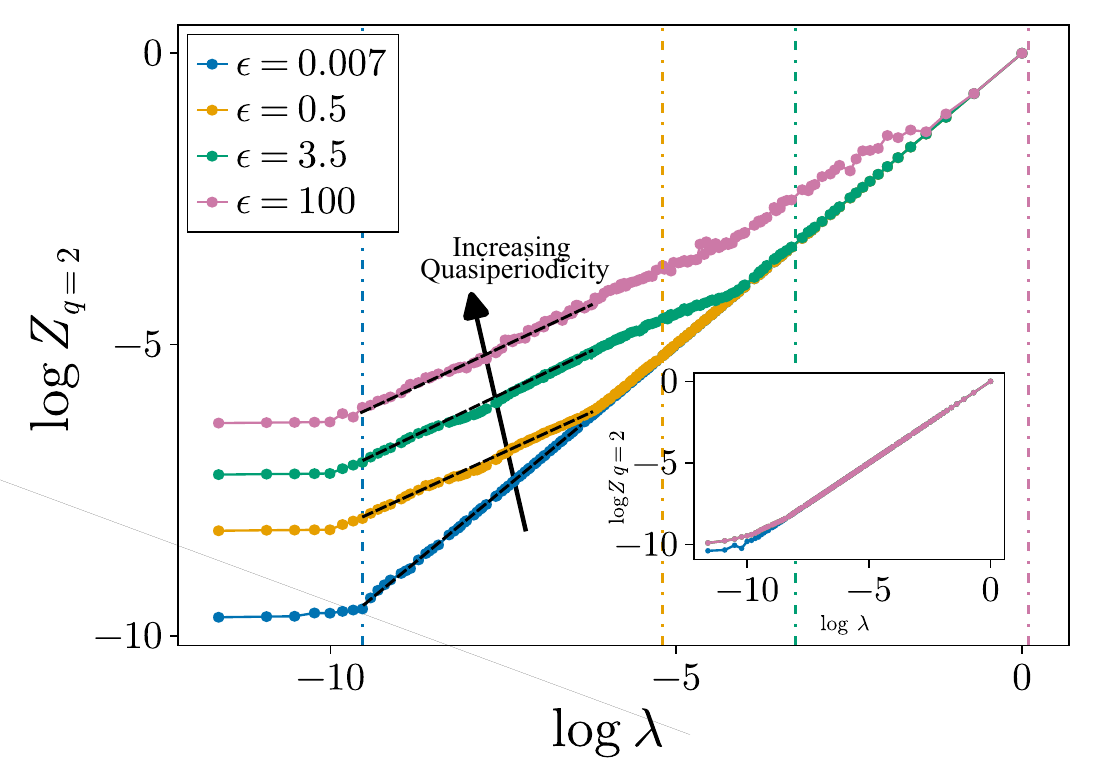}
    \caption{
    Partition function $ Z_{q = 2} $ as a function of the coarse-graining scale $ \lambda $ in the critical narrow bands, for different quasiperiodic approximants $( L \simeq 1.2\times10^{5} $). The curves range from the most periodic (blue) to the most quasiperiodic (magenta) case. Vertical dashed lines mark the scale corresponding to each unit cell. The slopes of the linear regions yield the multifractal exponent $ \tau_{q} $. Inset: same analysis for energies in the extended narrow bands, where all curves collapse with unit slope, consistent with Bloch-like behavior.
    }
    \label{fig:Z_lambda}
\end{figure}

All multifractal quantities of interest can be extracted through a scaling analysis of suitable functions with respect to  $\lambda$, as shown in Eqs.\,\eqref{eq:Power-law-scaling-for-Z} and \eqref{eq:Alpha-and-f}. In practice, this involves computing the partition function and entropy, analyzing their scaling under coarse-graining, and determining the corresponding exponents. Fig.\,~\ref{fig:Z_lambda} illustrates this procedure for the LDOS of our model, showing a log–log plot of $Z_{q=2}$ versus $\lambda$. The slope of each curve characterizes the LDOS behavior across length scales. A constant slope indicates power-law, scale-invariant behavior. At coarse resolution (large $\lambda$), the LDOS appears extended, with slope approaching 1. At finer scales (small $\lambda$), structural details emerge. In the quasiperiodic limit, scaling with multifractal exponents between 0 and 1 within the critical band signals multifractality, while in the extended narrow band, as shown in the inset, the slope remains unity, confirming extended states. From these slopes, we extract the multifractal exponents $\tau_q$, Hölder exponents $\alpha$, and corresponding Hausdorff dimensions $f(\alpha)$ shown in Fig.~\ref{fig:Singularity spectra}, through a fitting procedure, a linear least square optimization problem.

\begin{figure}[h!]
    \centering
    \includegraphics[width=0.98\linewidth]{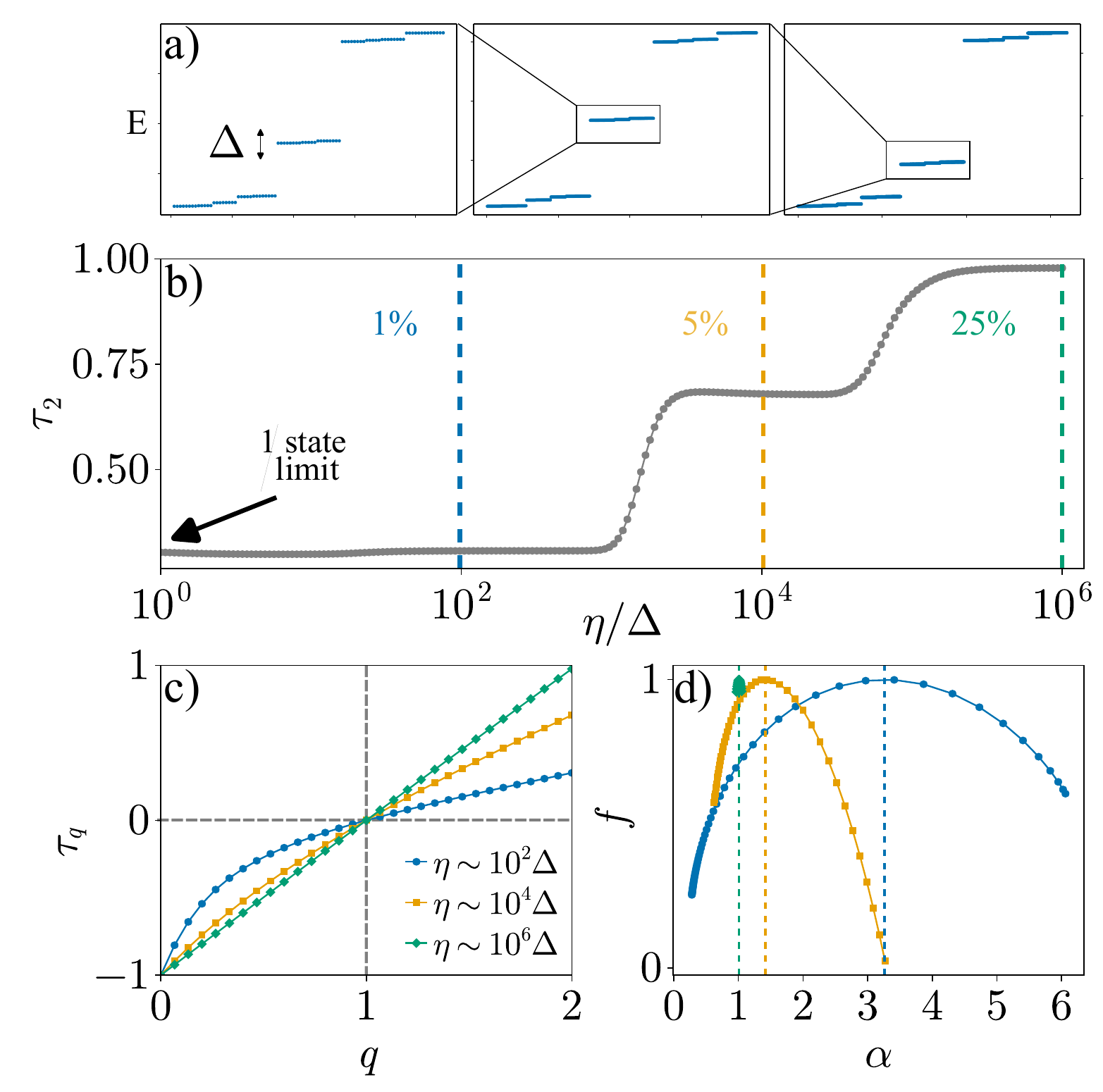}
    \caption{(a) Fractal structure of the spectrum. The characteristic energy scale $\Delta$ corresponds to the mean level spacing of the smallest miniband for this system size and is used as our reference energy scale. The rightmost zoom shows the same energy window highlighted in Fig.~1(d). (b) Multifractal exponent $\tau_2$ as a function of the LDOS broadening $\eta$, normalized by $\Delta$, for a system of size $L = 10946$. Vertical dashed lines mark the broadening values analyzed in detail, corresponding to $\sim$1\%, 5\%, and 25\% of states effectively contributing to the LDOS.
    (c) Multifractal exponents $\tau_q$ for representative broadening values $\eta/\Delta \sim 10^2$, $10^4$, and $10^6$.
    (c) Corresponding singularity spectra $ f(\alpha) $, showing the crossover from narrow to broad multifractal behavior as the energy resolution increases (i.e., as $\eta$ decreases).
    }
    \label{fig:Broadening}
\end{figure}

A natural question concerns the robustness of multifractality against energy broadening when computing the LDOS, a particularly relevant issue in STM experiments, where the energy resolution depends greatly on the temperature and the specific experimental apparatus. We investigate how both the multifractal exponent and the singularity spectrum of the LDOS evolve with increasing energy broadening. Our system features a fractal energy spectrum, characteristic of one-dimensional quasiperiodic models, with a Cantor-like structure. We define the reference energy scale $\Delta$ as the mean level spacing of the smallest finite-size miniband within the critical narrow band. This narrow band splits hierarchically into minibands and sub-minibands, up to the limit imposed by the finite system size. All eigenstates and eigenvalues were obtained via exact diagonalization.

As shown in Fig.\,\ref{fig:Broadening}(a), the energy spectrum exhibits a clear fractal hierarchy, where the characteristic energy scale $\Delta$ marks the smallest miniband mean level spacing. Fig.\,\ref{fig:Broadening}(b) displays the multifractal exponent $\tau_{q=2}$ increasing with the LDOS broadening $\eta$, exhibiting several plateaus. The jumps between plateaus correspond to the Gaussian broadening crossing the energy gaps separating minibands. When a sufficiently large fraction of states within a miniband is encompassed by the Gaussian, multifractality is eventually suppressed due to effective averaging over multiple states, yielding an LDOS with extended features. However, this occurs only when $\eta$ exceeds the characteristic energy scale $\Delta$ by several orders of magnitude. Panels \ref{fig:Broadening}(c) and \ref{fig:Broadening}(d) show the multifractal exponent and singularity spectrum for representative broadening values corresponding to distinct plateau regions in panel \ref{fig:Broadening}(b), where roughly 1\%, 5\%, and 25\% of the set of critical states in the model's spectrum contribute to the LDOS. The nonlinear behavior of $\tau_q$ and the broad $f(\alpha)$ curves persist even for large broadening values, demonstrating that, even in a system with a highly fractal spectrum and extremely small characteristic energy scales, multifractal signatures in the LDOS remain remarkably robust.

We emphasize that our choice of a 1D model, with its characteristic fractal energy spectrum, serves as a particularly stringent test case for the method's robustness. This scenario presents a significant challenge, not expected in 2D moir\'e materials. Here, any broadening risks averaging over the complex hierarchy of mini-bands, easily obscuring the underlying multifractal signatures. However, as our results demonstrate, the key results are resilient, persisting even when $\eta$ is large enough to encompass many of the fine spectral features. This gives us confidence in the method's applicability to 2D quasiperiodic systems, where the spectral structure is less complex and the current experimental capabilities already provide the high energy resolution LDOS maps required for this analysis. In practice, standard STM experiments have an energy resolution of $1$ meV, well within the moir\'e flat-bands, whilst novel state-of-art techniques break this barrier pushing this resolution to the order of $\mu$eV \cite{Fernandez-Lomana2021-yr,Odobesko2024}.

This approach is particularly promising for emerging quasiperiodic systems including i) moir\'e and super-moir\'e superlatices in twisted van der Walls heterostructures where quasiperiodic modulations naturally arise and ii) artifically structured systems where the quasiperiodic modulations can be engineered and the local density of states may be measured \cite{PhysRevLett.107.167404, Ter_Huurne2023-jl, PhysRevB.77.104201, Della_Villa2006-cs,PhysRevA.84.023831}.

This framework provides experimentalists with a model-independent diagnostic tool for identifying and characterizing quasiperiodic effects in systems where they might be overlooked in standard analysis approaches.

\begin{acknowledgments}
The authors acknowledge FCT-Portugal through Grant No. UIDB/04650/2020 is acknowledged. N.S. acknowledges support from FCT-Portugal through Grant No. 2024.00747.BD.
R.O acknowledges support from FCT-Portugal through Grant No. 2023.01313.BD.
\end{acknowledgments}

\bibliography{main}

\end{document}